\documentclass[11pt]{article}

\usepackage{setspace}
\usepackage[margin=1.25in]{geometry}

\usepackage[english]{babel}
\usepackage{amsmath}
\usepackage{amsfonts}
\usepackage{amssymb}
\usepackage{amsthm}
\usepackage{graphicx}
\usepackage{tikz-cd}
\usepackage{mathrsfs}
\usepackage{xfrac}
\usepackage{url}
\usepackage{color}
\usepackage[colorlinks=true, citecolor=blue]{hyperref}
\usepackage{enumitem}
\usepackage[usestackEOL]{stackengine}
\usepackage{accents}
\usepackage{bbm}
\usepackage{authblk}

\numberwithin{equation}{section}
\def\a{\alpha}

\def\be{\begin{equation}}
\def\ee{\end{equation}}
\def\ba#1\ea{\begin{align}#1\end{align}}
\def\no{\nonumber\\ }
\def\ra{\rangle}
\def\la{\langle}

\def\pexp{\mathscr{P}\!\exp}

\def\dom{\text{Dom}}
\def\pexp{\mathscr{P}\!\exp}

\newcommand{\ca}[1]{\mathcal{#1}}
\newcommand{\bb}[1]{\mathbb{#1}}
\newcommand{\fr}[1]{\mathfrak{#1}}
\newcommand{\scr}[1]{\mathscr{#1}}

\newcommand{\wt}[1]{\widetilde{#1}}

\newcommand{\ol}[1]{\overline{#1}}

\newtheorem{definition}{Definition}
\newtheorem{theorem}{Theorem}

\graphicspath{{Figures/}}

\title{{\huge Perfect quantum reflection from a cliff: \\a potential unbounded from below
\vspace{1cm}}}
\date{September 15, 2026}
\author[1,2]{Rodrigo Andrade e Silva\thanks{andradeesilvarodrigo@gmail.com}}
\affil[1]{\it \small Perimeter Institute for Theoretical Physics, Waterloo, ON, Canada}
\affil[2]{\it \small Maryland Center for Fundamental Physics, University of Maryland, College Park, MD, USA}

\begin{document}
\begin{titlepage}
\maketitle
\thispagestyle{empty}

\begin{abstract}
We explain how a potential that is well-defined everywhere on the positive half-line, but is nowhere positive and diverges to $-\infty$ as $x\rightarrow 0^+$, can nevertheless confine a particle to the half-line and lead to well-defined dynamics. Such perfect reflection from a ``cliff'' potential is achieved by a purely dynamical mechanism, without the need to impose any boundary conditions at the edge. We discuss in detail the role of self-adjointness in ensuring dynamical closure at the quantum level, and advocate the principle that, when no boundary conditions or other physical data are given, the formal quantized Hamiltonian is most naturally defined on its maximal domain. We then construct an explicit cliff potential with the claimed properties, showing that the Hamiltonian is self-adjoint on this maximal domain and therefore defines a dynamically closed quantum system. Finally, we study its energy eigenstates, spectrum, and the phase shift associated with the reflection.
\end{abstract}

\end{titlepage}

\tableofcontents
\vskip 3em

%%%%%%%%%%%%%%%%

\section{Introduction}
\label{sec:intro}

More than a century after its inception, quantum mechanics continues to surprise us with phenomena that defy our classical intuition. 
The famous quantum tunneling phenomenon is one such surprise encountered in the early days. Namely, a potential barrier which would be impenetrable for a classical particle, with a certain energy budget, suddenly becomes transparent for a quantum particle described by a wavefunction. Moreover, on the half-line, a ``wall'' potential that diverges to $+\infty$ as $x\to 0^+$ is completely impassable for classical particles, but quantum tunneling can still take place if the wall is ``sufficiently thin'', allowing a quantum particle to ``leak outside'' of the half-line. 
This may suggest that it is harder to confine a quantum particle to its configuration space than a classical particle. 
But that intuition fails, as we describe here a ``cliff'' potential that is nowhere positive and diverges to $-\infty$ as $x\rightarrow 0^+$, which would impose absolutely no obstruction to a classical particle, but happens to be impassable for a quantum particle. 

Examples of this phenomenon, where a system is dynamically open at the classical level but closed quantum-mechanically, have been observed before \cite{rauch1973two,reed1975ii}, mostly focusing on the behavior at infinity.
Our purpose is to develop a particularly explicit realization on the half-line, $\bb R^+ = (0, \infty)$, focusing on what happens at the edge, $x\to 0^+$, and to explore its physical, spectral and scattering properties.
We are also motivated by two broader questions.
First, we are interested in better understanding quantization in configuration spaces that lack a global linear structure, either due to having edges, non-trivial topology or being metrically incomplete, and here in particular focusing on aspects of time evolution. The half-line provides one of the simplest examples of a metrically-incomplete configuration space, which actually occurs in some situations of physical interest. 
For example, in minisuperspace models of cosmology \cite{Vilenkin_1994,pinto2013quantum,pinto2012wheeler,Kim_1997} (where the configuration variable is the volume of the universe) or in (1+1)-dimensional gravity \cite{harlow2020factorization} (as the only gauge-invariant property of a metric on a 1-dimensional space is the total proper length).
Second, we wish to clarify the important role of self-adjointness in ensuring dynamical closure at the quantum level, and its relation to quantization. Self-adjointness is sometimes confused with symmetricity (or Hermiticity), but those concepts are subtly distinct in infinite-dimensional Hilbert spaces. Upon quantization of a classical Hamiltonian, one usually obtains only a formal expression for the quantum Hamiltonian operator, but whether it is self-adjoint cannot be determined until a domain is specified. We advocate here the principle that, without additional physical data (such as boundary conditions), the Hamiltonian is most naturally defined on its maximal domain, i.e., the largest set of states on which the operator can act to produce other states in the Hilbert space. In this manner, the resulting operator may or may not be self-adjoint, despite its formal expression, and that is a physical implication of the quantization. 

We begin the paper with a pedagogical discussion of time evolution in quantum mechanics, focusing on the case of a particle living intrinsically on the half-line (Sec.~\ref{sec:timeevol}). 
We review the pertinent definitions surrounding the notion of self-adjointness, and provide a powerful criterion for determining whether a symmetric operator is actually self-adjoint (Sec.~\ref{subsec:selfadj}).
We then apply this machinery to the simple case of a free particle (Sec.~\ref{subsec:free}), illustrating how different domain choices can lead to different dynamical properties. 
This motivates our viewpoint that, a priori, the quantized Hamiltonian should be considered on its maximal domain.  Unsurprisingly, just as the classical particle can reach the edge of the half-line in finite time, the quantum Hamiltonian is not self-adjoint on its maximal domain and in fact does not generate well-defined time evolution. We also explain how the Hamiltonian can be made self-adjoint by suitable restrictions of the domain, but doing so requires additional physical data. 
We then review the limit point/limit circle criterion for self-adjointness (Sec.~\ref{subsec:limitpointcircle}), which will be employed when designing the cliff potential.

The core of the paper is the construction of the perfect reflecting cliff potential (Sec.~\ref{sec:potential}), which is well-defined everywhere on the half-line, nowhere positive, and diverging to $-\infty$ at the edge. We refer to it as the ``impassable cliff''. For simplicity, the potential is engineered as a series of piecewise-constant, self-repeating blocks, but its discontinuity is not essential for the mechanism. The crucial property of the potential is that it drives one of the solutions of the time-independent Schr\"odinger equation to divergence as $x\to 0^+$, ending up with infinite $L^2$ norm. 
We then discuss several dynamical properties of the cliff potential (Sec.~\ref{sec:dynamics}). 
In particular, we derive the energy eigenfunctions (Sec.~\ref{subsec:eigen}) by suitably determining initial conditions for the eigenvalue problem so as to prevent $L^2$ divergences as $x\to 0^+$.
The spectrum can be analyzed rather explicitly (Sec.~\ref{subsec:spectrum}). We show that it contains all non-negative energies (associated with scattering states) and a discrete negative portion (associated with bound states) that is unbounded from below. The negative spectrum has an interesting property of being approximately invariant under multiplication by $4$, at large negative energies.
Finally, we compute the phase shift acquired upon reflection from the cliff potential (Sec.~\ref{subsec:phase}), in a scattering scenario.
We end with a brief conclusion (Sec.~\ref{sec:conc}), reiterating the main aspects of the paper.

\section{Time evolution on the half-line}
\label{sec:timeevol}

Start with a classical particle of mass $m$ on the half-line, $\bb R^+ = (0,\infty)$, described by the Hamiltonian
\be
H = \frac{p^2}{2m} + V(x)
\ee
where $V : \bb R^+ \to \bb R$ is a real-valued, piecewise continuous potential. 
The specification of the potential $V$ is sufficient to determine the time evolution of particle, at least for an interval of time while the particle remains in its domain $\bb R^+$. In particular, the time evolution is well-defined for all times as long as $V$ is unbounded from above as $x \to 0^+$ and it does not decrease too fast as $x\to +\infty$.\footnote{\samepage If the potential is unbounded from above for $x \ge x_0 >0$, the particle will never escape towards infinity. On the other hand, if the potential is bounded from above for $x \ge x_0 >0$, it may occur that it can escape to infinity in finite time. If at $x=x_0$ it has positive velocity, then its velocity at position $x \ge x_0$ will be $v = \sqrt{2m(H - V)}$, provided that $H > V(x)$. The time to reach infinity is 
\[
T = \int_{x_0}^\infty \frac{dx}{\sqrt{2m(H - V)}} \,,
\]
and hence the dynamics is well-defined for all time as long as $T = \infty$ for all $H > V$. For example, if $V(x) = -\kappa x^\alpha$, with $\kappa > 0$, the integral diverges for all $H > V$ if and only if $\alpha \le 2$.}
In that case, the system is said to be \emph{classically closed}.

Now consider the quantization of that system, based on the Hilbert space 
\be
\ca H = L^2(\bb R^+)\,.
\ee
Formally, the Hamiltonian is realized as the operator
\be\label{Hquant}
H = -\frac{\hbar^2}{2m}\frac{d^2}{dx^2} + V(x)
\ee
However, an operator is not completely defined until its domain, $\ca D := \dom(H) \subset \ca H$, is specified. The minimal requirement for the domain is that for every $\psi \in \ca D$, $H\psi \in \ca H$. It is also standard to assume that $\ca D$ is dense in $\ca H$, so there is no state orthogonal to all states on which $H$ can act. While a bounded operator can always be defined on the entire Hilbert space, an unbounded operator like $H$ must always be defined on a proper subspace.
This brings the question: 
\[
\text{\emph{How to select the domain $\ca D$ of the Hamiltonian?}}
\]
A priori, a very natural choice, in the absence of further physical data, is to take $\ca D$ to be the \emph{maximal domain}, $\ca D_\text{max}$. That is, the largest subspace of $\ca H$ on which $H$ is well-defined. More explicitly,
\be\label{Dmax}
\ca D_\text{max} = \big\{ \psi \in L^2(\bb R^+),\,\text{such that } H\psi \in L^2(\bb R^+) \big\}
\ee
The second derivative (inside $H$) acting on $\psi$ must be understood distributionally.\footnote{\samepage That is, for any compactly-supported smooth function $f$,
\[
\int_0^\infty\!dx\, \frac{d^2\psi}{dx^2} f(x) = \int_0^\infty\!dx\, \psi(x)\frac{d^2f}{dx^2} 
\]
This ensures that the operator captures correctly any ``kinks'' in a wavefunction. For example, if $\psi$ is given by $|x-x_0|$ in some neighborhood of $x_0$, its second derivative vanishes almost everywhere in that neighborhood, but distributionally $\psi''(x) = 2\delta(x-x_0)$, which is not square integrable and consequently such a $\psi$ is excluded from $\ca D_\text{max}$.\label{distrder}}
Another reasonable choice, a priori, is what we call the \emph{minimal domain}, $\ca D_\text{min}$. By that we mean the domain consisting of compactly-supported smooth functions,
\be
\ca D_\text{min} = C^\infty_c(\bb R^+)
\ee
The nomenclature in this case may be misleading, since there are even smaller dense domains---we just mean that this is a very simple, natural and ``small'' domain. Later we will also consider other domains produced by restrictions and extensions.

The time evolution of states with respect to the Hamiltonian is given by the Schr\"odinger equation
\be
i\hbar \frac{d\psi(t)}{dt} = H\psi(t)
\ee
Formally, the solution is given by the exponential operator
\be
\psi(t) = e^{-iHt/\hbar}\psi(0)
\ee
The system is said to be \emph{quantum-mechanically closed} if $e^{-iHt/\hbar}$ is well-defined and unitary. In that way, a state in $\ca H$ will evolve into another state in $\ca H$ with the same norm. Physically, this ensures that time evolution is well-defined, deterministic and that the particle remains on the configuration space with $100\%$ probability for all times.
The question is then:
\[
\text{\emph{Is $e^{-iHt/\hbar}$ well-defined and unitary?}}
\]
And the answer, according to Stone's theorem, is:
\[
\text{\emph{If and only if $H$ is self-adjoint.}}
\]
At first sight, it may seem that $H$ is self-adjoint for any real-valued potential $V(x)$, since the second derivative term is $P^2/2m$, and the momentum $P$ is at least formally self-adjoint. 
But, in fact, in infinite dimensional Hilbert spaces the concept of self-adjointness is more subtle and depends on the precise definition of the domain.

\subsection{Self-adjointness}
\label{subsec:selfadj}

In this subsection we will review the pertinent definitions underlying the concept of self-adjointness, and state a useful theorem that gives a criterion for an operator to be self-adjoint.
For standard references see, for example, \cite{reed1975ii,hall2013quantum}, and for some shorter pedagogical notes see \cite{bonneau2001self,araujo2004operator}. 

\begin{definition}[Adjoint] 
Let $A$ be a (possibly unbounded) linear operator from a dense subspace $\dom(A) \subset \ca H$ into $\ca H$. We define the adjoint of $A$, denoted by $A^\dag$, as the linear operator from some domain $\dom(A^\dag) \subset \ca H$ into $\ca H$ such that
\be\label{Adj}
\la \psi | A^\dag \phi \ra = \la A \psi | \phi \ra 
\ee
for all $\psi \in \dom(A)$. The domain of $A^\dag$ is defined as the set of all $\phi \in \ca H$ such that a state $A^\dag\phi$ with the property above exists. 
\end{definition}
By Riesz theorem, a state $\chi = A^\dag\phi$ satisfying \eqref{Adj} exists if and only if the functional $\psi \mapsto \la \phi | A\psi \ra$ is bounded, i.e., if there exists a constant $C$ such that $|\la \phi | A\psi \ra | \le C |\!| \psi |\!|$ for all $\psi \in \dom(A)$. The ``only if'' direction is easy to see because $|\la \phi | A\psi \ra | = |\la \chi | \psi \ra | \le |\!|\chi|\!|\, |\!| \psi |\!|$, so we can take $C = |\!|\chi|\!|$.
\begin{definition}[Symmetricity]
The operator $A$ is said to be symmetric if $A = A^\dag$ on $\dom(A)$ and $\dom(A) \subset \dom(A^\dag)$. 
\end{definition}
\begin{definition}[Self-adjointness]
The operator $A$ is said to be \emph{self-adjoint} if it is symmetric and $\dom(A) =\dom(A^\dag)$. 
\end{definition}
Since the specification of the domain of an operator is a fundamental part of its definition, we can think of self-adjointness as the condition that the operator is \emph{equal} to its adjoint. Note that a symmetric operator is not necessarily self-adjoint because it may occur that $\dom(A) \subsetneq \dom(A^\dag)$. This inclusion property can be seen by observing that, if $A$ is symmetric, then for any $\psi \in \dom(A)$ we have that $A^\dag \psi = A\psi$ satisfies \eqref{Adj}, so $\psi \in \dom (A^\dag)$; on the other hand, there can be states outside of $\dom(A)$ for which $A^\dag\psi$ exists. The relevant concept for the \emph{spectral theorem} is self-adjointness, not symmetricity.\footnote{The spectral theorem holds more generally for \emph{normal operators}, which are densely-defined operators $N$ such that $N N^\dag = N^\dag N$ with $\dom(N N^\dag) = \dom(N^\dag N)$. The main difference is that the spectrum of $N$ is, in general, contained in $\bb C$ rather than $\bb R$.}

An important point to notice is that if $A$ is densely-defined, its adjoint is always \emph{closed}. That is, the graph of $A^\dag$, $(\phi, A^\dag\phi)$ with $\phi \in \dom(A^\dag)$, is a closed subset of $\ca H \oplus \ca H$. Thus, $A$ can only be self-adjoint if it is closed. This leads to the following concept: 
\begin{definition}[Essential self-adjointness]
A densely-defined operator $A$ is said to be essentially self-adjoint if it is closable and its closure is self-adjoint. (An operator is said to be closable if the closure of its graph in $\ca H \oplus \ca H$ is the graph of a function, and this function defines its closure.) 
\end{definition}
If $A$ is essentially self-adjoint, then its closure is the \emph{unique} self-adjoint extension of $A$.
(An extension of $A$ is an operator $\wt A$ with $\dom(\wt A) \supset \dom(A)$ which equals $A$ on $\dom(A)$.)
Then, given a domain in which the operator is essentially self-adjoint, we can uniquely extend it into a self-adjoint operator, if for physical reasons that is desirable.

The definition of self-adjointness above suggests that it may be delicate to determine whether a given operator is self-adjoint. Namely, one would in principle have to evaluate the domain of its adjoint and check whether that matches with its own domain.
Although this is feasible in many cases, there are some useful theorems which provide convenient criteria for checking self-adjointness.
We now state an important one, based on \emph{deficiency indices}.
\begin{definition}[Deficiency indices]
Let $A$ be a symmetric operator and consider the subspaces
\ba
\ca K_+ &= \ker(A^\dag - i) \\
\ca K_- &= \ker(A^\dag + i)
\ea
The numbers $n_+ := \dim(\ca K_+)$ and $n_- := \dim(\ca K_-)$ are called the ``deficiency indices'' of $A$.  
\end{definition}
In words, $n_+$ and $n_-$ are the dimensions of the eigenspaces of $A^\dag$ associated with the eigenvalues $+i$ and $-i$, respectively.
We have:
\begin{theorem}\label{TheoSAdefic}
Let $A$ be a closed symmetric operator with deficiency indices $n_+$ and $n_-$. Then, $A$ is self-adjoint if and only if $n_+ = n_- = 0$. Moreover, $A$ admits self-adjoint extensions if and only if $n_+ = n_-$. There is a one-to-one correspondence between self-adjoint extensions of $A$ and unitary maps from $\ca K_+$ onto $\ca K_-$.
\end{theorem}
Notice that the ``only if'' part of the condition for $A$ to be self-adjoint is quite straightforward: the spectrum of a self-adjoint operator must be contained in the real line, so a self-adjoint operator cannot have eigenvalues $\pm i$. 
What is remarkable is that the condition is also sufficient: a (closed) symmetric operator whose adjoint does not have eigenvalues $\pm i$ must be self-adjoint. 
Since a symmetric operator can only have real eigenvalues, we see that when $A$ is symmetric but fails to be self-adjoint, then $A^\dag$ is necessarily not symmetric.

\subsection{An example: the free particle}
\label{subsec:free}

To gain intuition on the distinction between symmetricity and self-adjointness, and the importance of self-adjointness for dynamical closure, let us study the simple case of a free particle ($V=0$) on $\bb R^+$. 
Classically, that system is evidently \emph{not} dynamically closed. A particle initially moving to left will reach $x=0$ in finite time, and time evolution becomes ill-defined.
It would not be surprising that, quantum mechanically, the Hamiltonian operator 
\be
H = -\frac{\hbar^2}{2m} \frac{d^2}{dx^2}
\ee
fails to be self-adjoint (and consequently $e^{-iHt/\hbar}$ is ill-defined or fails to be unitary).

Assume initially that $H$ is defined on its maximal domain, $\ca D_\text{max}$, introduced previously. According to footnote~\ref{distrder}, since the derivative acts distributionally on wavefunctions, we must exclude from the domain any wavefunctions which are discontinuous or have ``kinks''. In fact, $\ca D_\text{max}$ is the Sobolev space $W^{2,2}(\bb R^+)$.\footnote{The \emph{Sobolev space} $W^{k,p}(U)$, where $k\in \bb N$, $1\le p \le \infty$ and $U$ is a measurable set, is defined as the space of all functions $f: U \rightarrow \bb R$ such that $f$ and (weak) derivatives of order equal or less than $k$ are in $L^p(U)$.} 
It turns out that $H$ is not even symmetric.
For $\psi \in \ca D_\text{max}$, upon integration by parts we have
\ba
\la H\psi| \phi\ra &= \int_0^\infty\!dx\left( -\frac{\hbar^2}{2m} \frac{d^2\psi^*}{dx^2}\right) \phi \no
&= \int_0^\infty\!dx\,\psi^*\left( -\frac{\hbar^2}{2m} \frac{d^2\phi}{dx^2}\right) - \frac{\hbar^2}{2m} \left(\frac{d\psi^*}{dx}\phi - \psi^* \frac{d\phi}{dx}\right)\!\bigg|_0 \label{freeHphipsi}
\ea
while for $\psi, \phi \in \ca D_\text{max}$,
\be
\la \psi| H\phi\ra = \int_0^\infty\!dx\,\psi^*\left( -\frac{\hbar^2}{2m} \frac{d^2\phi}{dx^2}\right)
\ee
and, since the boundary terms in \eqref{freeHphipsi} do not necessarily vanish, $\la H\psi| \phi\ra \ne \la \psi| H\phi\ra$. 
Consequently, $H$ cannot be self-adjoint with its maximal domain.

It is possible, nonetheless, to restrict the domain of the Hamiltonian so that it is symmetric. In particular, let us consider restrictions that produce a \emph{maximal symmetric} operator.\footnote{A ``maximal symmetric'' operator is a closed symmetric operator with no non-trivial symmetric extensions.} Let $\ca D_\text{max}^\text{sym} \subsetneq \ca D_\text{max}$ be the domain of one such restriction. 
In order to remove the boundary terms in \eqref{freeHphipsi}, $\ca D_\text{max}^\text{sym}$ must be a subset of $\ca D_\text{max}$ satisfying, at very least, the boundary condition
\be\label{DmaxsymBC}
\left(\frac{d\psi^*}{dx}\phi - \psi^* \frac{d\phi}{dx}\right)\!\bigg|_0 = 0
\ee
for $\psi,\phi \in \ca D_\text{max}^\text{sym}$.
Viewing this condition as the vanishing of the cross product between vectors $(\psi(0)^*, \phi(0))$ and $(\psi'(0)^*, \phi'(0))$, we see that $(\psi'(0)^*, \phi'(0)) \propto (\psi(0)^*, \phi(0))$. Since the relation satisfied by $\psi$ must be the same as the one satisfied by $\phi$, and they must be satisfied independently, we note that the proportionality factor must be a real constant. Thus, with $\lambda \in \bb R \cup \{\infty\}$, we have
\be
\frac{d\psi}{dx}(0) = \lambda \psi(0)
\ee
where $\lambda = \infty$ is understood as the Dirichlet boundary condition, $\psi(0) = 0$, and $\lambda = 0$ corresponds to the Neumann boundary condition, $\psi'(0) = 0$.
In this manner, $\ca D_\text{max}^\text{sym}$ can be labeled by the parameter $\lambda \in \bb R \cup \{\infty\}$,
\be
\ca D_\text{max}^\text{sym}(\lambda) =  \Big\{ \psi \in L^2(\bb R^+),\,\text{such that } H\psi \in L^2(\bb R^+) \text{ and } \frac{d\psi}{dx}(0) = \lambda \psi(0) \Big\}
\ee
Let us denote by $H_\lambda$ the Hamiltonian defined on $\ca D_\text{max}^\text{sym}(\lambda)$. Since $H_\lambda$ is symmetric, we can check whether it is self-adjoint. 
From \eqref{freeHphipsi}, with $\psi \in \ca D_\text{max}^\text{sym}(\lambda)$, the adjoint $H_\lambda^\dag$ satisfies, for $\lambda \ne \infty$,
\be
\la \psi| H_\lambda^\dag\phi\ra = \la H_\lambda\psi | \phi\ra = \int_0^\infty\!dx\,\psi^*\left( -\frac{\hbar^2}{2m} \frac{d^2\phi}{dx^2}\right) - \frac{\hbar^2}{2m} \psi^*(0) \left(\lambda\phi(0) -\frac{d\phi}{dx}(0)\right)
\ee
In order for a state $H_\lambda^\dag\phi$ satisfying this equation to exist, the boundary terms must vanish. Since $\psi^*(0)$ is free, this implies $\phi'(0) = \lambda \phi(0)$. If $\lambda = \infty$,
\be
\la \psi| H_\infty^\dag\phi\ra = \la H_\infty \psi|\phi\ra = \int_0^\infty\!dx\,\psi^*\left( -\frac{\hbar^2}{2m} \frac{d^2\phi}{dx^2}\right) - \frac{\hbar^2}{2m} \frac{d\psi^*}{dx}(0) \,\phi(0)
\ee
and since now $\psi'(0)$ is free, we need $\phi(0) = 0$. Thus, for every $\lambda \in \bb R\cup\{\infty\}$, the boundary conditions satisfied by wavefunctions in the domain of $H_\lambda^\dag$ are the same as those defining the domain of $H_\lambda$. That is,
\be
\dom(H_\lambda^\dag) = \ca D_\text{max}^\text{sym}(\lambda) = \dom(H_\lambda)
\ee
revealing that $H_\lambda$ is self-adjoint.

Alternatively, we could have considered the Hamiltonian defined on its minimal domain, $\ca D_\text{min}$, which we denote by $H_\text{min}$. With this domain, $H_\text{min}$ is not closed, but as we mentioned before we can always consider its closure, $\ol H_\text{min}$, whose domain we denote by $\ol{\ca D}_\text{min}$. It turns out that
\be
\ol{\ca D}_\text{min} = \Big\{ \psi \in L^2(\bb R^+),\,\text{such that } H\psi \in L^2(\bb R^+) \text{ and } \psi(0) = \frac{d\psi}{dx}(0) = 0 \Big\}
\ee
Because the boundary terms in \eqref{freeHphipsi} vanish in $\ol{\ca D}_\text{min}$, $\ol H_\text{min}$ is symmetric.\footnote{Note that since $\ol{\ca D}_\text{min} \subsetneq \ca D_\text{max}^\text{sym}(\lambda)$, for every $\lambda \in \bb R \cup \{\infty\}$, $\ol H_\text{min}$ admits non-trivial symmetric extensions, $H_\lambda$, and hence it is not maximal symmetric.} In fact, since the boundary terms vanish given only that $\phi \in \ol{\ca D}_\text{min}$, without any restrictions on $\psi \in \dom(\ol H_\text{min}^\dag)$, we conclude that $\ol H_\text{min}^\dag$ is defined on the maximal domain,
\be
\dom(\ol H_\text{min}^\dag) = \ca D_\text{max}
\ee
Since $\dom(\ol H_\text{min}) \subsetneq \dom(\ol H_\text{min}^\dag)$, the Hamiltonian is \emph{not} essentially self-adjoint on its minimal domain. 
In fact, this can be confirmed from Theorem~\ref{TheoSAdefic} by evaluating the deficiency indices. The eigensubspace of $\ol H_\text{min}^\dag$ associated with $+i$ is defined by the equation
\be
\ol H_\text{min}^\dag \phi = i \phi
\ee
The solutions are
\be
\phi \propto a e^{-\sqrt{m}(1-i)x/\hbar} + b e^{\sqrt{m}(1-i)x/\hbar}
\ee
The solutions with $b\ne 0$ are not square-integrable, but if $b=0$ the solutions belong to the domain of $\ol H_\text{min}^\dag$, $\ca D_\text{max}$. The eigenspace associated with $+i$ is thus 1-dimensional, so the corresponding deficiency index is $n_+ = 1$. Similarly one can verify that $n_- = 1$. This confirms that $\ol H_\text{min}$ is not self-adjoint, but since $n_+ = n_-$ it has self-adjoint extensions. In fact, we know that these extensions are produced by extending $\ol{\ca D}_\text{min}$ into $\ca D_\text{max}^\text{sym}(\lambda)$, for any $\lambda \in \bb R \cup \{\infty\}$. 
We can also understand the one-to-one correspondence between self-adjoint extensions and unitary maps from $\ca K_+$ onto $\ca K_-$ stated in the theorem: since $\ca K_\pm$ are 1-dimensional, the unitary maps are labeled by a phase $e^{i\eta}$, so we can take $\eta \in \bb R/2\pi\bb Z$, which is isomorphic to $S^1$ just like $\bb R\cup \{\infty\}$.

Let us discuss the physical interpretation of the analysis above.
First, as we explained, there is no a priori reason that the Hamiltonian defined on an arbitrary domain would be self-adjoint. Classically, the system is not dynamically closed, so it would not be unreasonable that its quantization is not quantum-mechanically closed. In fact, that is the natural outcome, since we showed that $H$ is not self-adjoint when defined on its natural maximal domain, without imposing any extra conditions.
Second, notice that the self-adjoint restrictions of the Hamiltonian are labeled by a parameter $\lambda \in \bb R\cup\{\infty\} \simeq S^1$. This parameter defines a boundary condition at $x=0$, and it has physical implications: it determines how a wavepacket ``reflects'' when it reaches $x=0$, with the phase shift depending on $\lambda$. Notice that $\lambda$ represents extra physical data, beyond the specification of the configuration space and the formal quantization of the classical Hamiltonian. The classical counterpart would be specifying what happens when the particle reaches $x=0$. For example, one could consider a situation where $x=0$ is not merely an open edge of space, but rather it is a ``hard wall'' which causes the particle to elastically reflect upon impact. This forces the classical system to close dynamically, in the same way that the suitable domain restriction closes the quantum system. One difference is that putting a reflecting boundary condition at $x=0$ is the only way to close the classical system while maintaining continuity of the trajectories and conserving energy, while in quantum mechanics there is a ``circle'' of possible dynamical closures, since the phase shift acquired upon reflection is physically meaningful.

A possible lesson we derive is the following.
\emph{When quantizing a classical system for which no boundary conditions are given, the formal quantized Hamiltonian is most naturally defined on its maximal domain. Whether this Hamiltonian is self-adjoint or not, and consequently whether the system is dynamically closed or not, is a consequence of the quantization. If classically there are boundary conditions at the edges of the configuration space, those conditions can be ``quantized'' by means of selecting corresponding domain restrictions of the Hamiltonian. Those domain restrictions may require additional physical data, beyond what is provided classically, and they may modify the status of the dynamical closure at the quantum level.} (The other natural choice, the closure of the minimal domain Hamiltonian, is effectively equivalent since it is self-adjoint if and only if the maximal domain Hamiltonian is self-adjoint---see footnote~\ref{HminHmax}.)

\subsection{The limit point/limit circle criterion}
\label{subsec:limitpointcircle}

To end this section, we present a final useful criterion for determining self-adjointness for potentials of the form given in \eqref{Hquant}, known as the limit point/limit circle criterion.

To motivate this criterion, let us understand why the free Hamiltonian, studied in the previous subsection, can be self-adjoint or not depending on each domain considered.
From the Schr\"odinger equation, we can define the \emph{probability current} of a wavefunction $\psi$ by 
\be
J_\psi(x) := \frac{\hbar}{2mi} \left[ \psi^*(x) \frac{d\psi}{dx}(x) -  \psi(x) \frac{d\psi^*}{dx}(x) \right]
\ee
where $x\in \bb R^+$.
If $J_\psi(x)$ can be non-zero as $x\to 0^+$, the wavefunction can ``exchange'' norm with the ``outside world''. This explains why $H$ is not self-adjoint on $\ca D_\text{max}$, as there is no restriction on $\psi(0)$ and $\psi'(0)$. 
Observe that the pathology is not merely that time evolution is non-unitary, but rather it is ill-defined. There are too few boundary conditions for making the Schr\"odinger equation well-posed as a boundary-initial value problem.
Intuitively, the evolution is not deterministic since one cannot know what is coming in from the ``outside'' through $x=0$. 
Notably, the condition \eqref{DmaxsymBC} necessary to make the Hamiltonian maximal symmetric precisely ensures that $J_\psi(x)$ vanishes as $x\to 0^+$. Thus, it is not surprising that the Hamiltonian turned out to be self-adjoint with domain $\ca D_\text{max}^\text{sym}(\lambda)$. 
On the other hand, for the closure of the minimal domain Hamiltonian, too many boundary conditions are imposed at $x=0$, namely $\psi(0) = \psi'(0) = 0$. While this also makes $J_\psi(x)$ vanish as $x\to 0^+$, these conditions are too strong, again making time evolution ill-posed. In this case, the Schr\"odinger equation would try to evolve initial states into states outside the domain of $\ol H_\text{min}$, at which point evolution cannot be continued. 

Take the general Hamiltonian given in \eqref{Hquant}, with an arbitrary piecewise continuous potential $V(x)$, defined on its maximal domain. Now look at the eigenvalue problem
\be\label{eigenproblem}
-\frac{\hbar^2}{2m}\frac{d^2\psi}{dx^2} + V\psi = E\psi
\ee
where $E \in \bb R$. As this differential equation is of second order, it must have two linearly-independent solutions, which we denote by $\zeta$ and $\varphi$. Say that both solutions have finite near-zero norm.\footnote{We say that a function $\psi : \bb R^+ \rightarrow \bb C$ has \emph{finite near-zero norm} if $\int_0^{\epsilon}\!dx\, |\psi(x)|^2 < \infty$ for some $\epsilon > 0$, and \emph{infinite near-zero norm} otherwise.}
As the equation has real coefficients, it is possible to choose a basis of real solutions, so we assume that $\zeta$ and $\varphi$ are taken to be real. Then, for a generic (complex) linear combination, 
\be
J_{\alpha \varphi + \beta \zeta} = \frac{\hbar}{2mi}(\alpha \beta^* - \beta \alpha^*) W(\varphi,\zeta)
\ee
where $W$ is the Wronskian, which is non-zero since the two solutions are linearly independent.
Therefore, generic locally admissible solutions carry nonzero probability current through the endpoint, indicating that the maximal domain Hamiltonian cannot be self-adjoint in this case.
Now suppose that $\zeta$ has infinite near-zero norm while $\varphi$ has finite near zero norm. Since \eqref{eigenproblem} is an equation with real coefficients, $\varphi^*$ is also a solution. As $\varphi^*$ also has finite near-zero norm, it must belong to the same 1-dimensional subspace spanned by $\varphi$, and hence $\varphi^* = c\varphi$ for some complex number $c$. 
This implies that the probability current $J_\varphi(x)$
vanishes everywhere. 
Notice that the near-zero norm behavior of the solutions to \eqref{eigenproblem} are determined by their divergence behavior near zero, which are not affected by the (finite) value of $E$. Therefore, for every $E$, one solution would behave like $\zeta$ near zero, which does not belong to the Hilbert space, and all eigenstates of $H$ would behave like $\varphi$ near zero. Therefore, the generic wavefunction in the domain of $H$ would be given by superpositions of $\varphi$-like functions near zero, which all have zero probability flux as $x\to 0^+$, indicating that $H$ can be self-adjoint in this case, as far as that endpoint is concerned.
For $H$ to be self-adjoint, a similar behavior must occur as $x\to\infty$, that is, at least one solution of \eqref{eigenproblem} must have infinite near-infinity norm.
The following theorem by Weyl \cite{Weyl1910,zettl2005sturm} establishes this result rigorously:
\begin{theorem}
Let $V$ be a piecewise continuous, real valued potential on $\bb R^+$. If all solutions of \eqref{eigenproblem} for some (and thus all) $E$ have finite near-zero norm, then $V$ is said to be in the ``limit circle case'' at zero; otherwise it said to be in the ``limit point case'' at zero. The same classification applies for infinity, based on the near-infinity norm of the solutions. Then $H$ is self-adjoint on its maximal domain, $\ca D_\text{\rm max}$, if and only if $V$ is in the limit point case at both zero and infinity.\footnote{The theorem stated in \cite{reed1975ii} has two differences. First, the potential is assumed continuous. But it suffices to assume that it is locally integrable, and in particular piecewise continuous. Second, instead of ``$H$ is self-adjoint on its maximal domain'' it says ``$H$ is essentially self-adjoint on its minimal domain''. The two formulations are equivalent, for the following reason. It is equivalent to say that $H$ is essentially self-adjoint on $\ca D_\text{min}$ and that $H$ is self-adjoint on $\ol{\ca D}_\text{min}$. Denote by $\ol H_\text{min}$ and $H_\text{max}$ the Hamiltonian defined with domain $\ol{\ca D}_\text{min}$ and $\ca D_\text{max}$, respectively. Under the assumed conditions for $V$, we have $\ol H_\text{min}^\dag = H_\text{max}$. Since $\ol H_\text{min}$ is closed and densely-defined, its double adjoint equals to itself, so $\ol H_\text{min} = H_\text{max}^\dag$. Therefore $\ol H_\text{min}^\dag = \ol H_\text{min} \Longleftrightarrow H_\text{max}^\dag = H_\text{max}$, which means $\ol H_\text{min}$ is self-adjoint if and only if $H_\text{max}$ is self-adjoint. Moreover, they are equal if they are self-adjoint.\label{HminHmax}}
\end{theorem}
As a simple application, note that for the free Hamiltonian, the eigenstates are linear combinations of $e^{\pm i\sqrt{2mE}x/\hbar}$. For any $E$, these have finite near-zero norm, so $V=0$ falls in the limit circle case and, consequently, the maximal domain Hamiltonian is not self-adjoint.

\section{The impassable cliff}
\label{sec:potential}

In this section we construct a potential $V: \bb R^+ \rightarrow \bb R$, diverging to $-\infty$ as $x\rightarrow 0^+$ and converging to zero as $x\rightarrow +\infty$, which is naturally confining to the half-line, without imposing any boundary conditions at $x=0$. That is, the corresponding Hamiltonian is self-adjoint with its maximal domain, and consequently time evolution is unitary in the Hilbert space $\ca H = L^2(\bb R^+)$. We will refer to this potential as the ``impassable cliff'', for its ability to prevent any quantum particle from passing through.

Before proceeding, it is worth comparing again dynamical closure in classical and quantum systems, in view of the limit point/limit circle criteria of the previous section. Classically, as discussed earlier, any potential bounded from above is dynamically open. Quantum mechanically, the only directly analogous result is that for a bounded potential (from both above and below), the system is also dynamically open. This can be seen directly from the limit circle/limit point criterion. Like in the case of $V=0$, a bounded potential cannot modify the eigenstates of the Hamiltonian enough so as to change their bounded behavior near zero. 
When the potential diverges near zero, dynamical closure can occur. In particular, Theorem X.10 of \cite{reed1975ii} establishes that if $V(x) \ge (3\hbar^2/8m)x^{-2}$ in a neighborhood of zero then it is in the limit point case at zero (thus dynamically closed). In quantum mechanics, however, the potential being unbounded from above near zero is not sufficient for ensuring dynamical closure: if $0< V(x) \le (3\hbar^2/8m - \epsilon)x^{-2}$ near zero, for some $\epsilon>0$, then it is in the limit circle case at zero (thus dynamically open). 
Fairly generally, the system will be quantum mechanically open if the potential diverges to $-\infty$ as $x\to 0^+$, like in the classical case. For example, a decreasing potential near zero is always in the limit circle case. 
Those results may give the impression that it is ``harder'' for a system to be quantum-mechanically closed.
However, the example of the impassable cliff shows that a potential trending towards $-\infty$ near zero, which is categorically open classically, can actually be quantum-mechanically closed.

Consider a Hamiltonian of the form \eqref{Hquant}, with $V(x) \le 0$, defined on its maximal domain $\ca D_\text{max}$ as defined in \eqref{Dmax}.
Let 
\be
k(x) := \frac{\sqrt{-2m V(x)}}{\hbar}
\ee
The $E=0$ eigenvalue problem then reads
\be\label{eigenEasymp}
\frac{d^2 \psi}{dx^2}(x) = -k(x)^2\psi(x)
\ee
Let $\zeta$ and $\varphi$ denote two linearly independent solutions to this equation.
The strategy is to construct a $k(x)$ that oscillates as $x \rightarrow 0^+$ in such a way as to create a ``resonance'' driving $\zeta$ to infinity in a finite distance. We would then expect that the other solution, $\varphi$, which is completely ``out of phase'' with the oscillations of the potential, would be driven to zero. 
If $\zeta$ has infinite near-zero norm, we would be in the limit point case and the Hamiltonian would then be self-adjoint.

\begin{figure}  
\centering
\includegraphics[scale = 0.62]{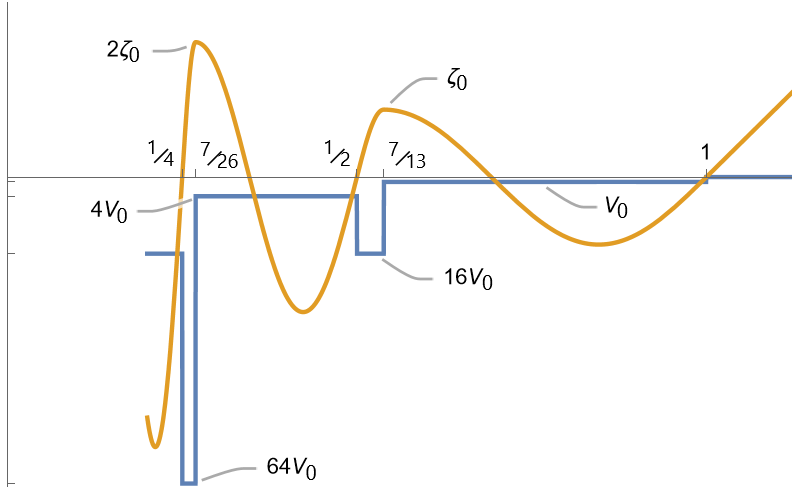}
\caption{\small The ``impassable cliff'' potential $V(x)$ is shown in blue. The edge is at $x_0 = 1$, so the cliff is in the region $(0,1)$, and $V_0 := -\hbar^2 k_0^2/2m$. Only the first two blocks are displayed entirely: the first is in $(\sfrac{1}{2},1)$ and the second is in $(\sfrac{1}{4}, \sfrac{1}{2})$. The first stage of the first block is in $(\sfrac{7}{13},1)$ and the second stage is in $(\sfrac{1}{2}, \sfrac{7}{13})$. The diverging solution $\zeta$ is shown in orange.}
\label{fig:potentialV}
\end{figure}

To produce an explicit example of such a potential, we go as follows.
Say that we start at $x = x_0$ with $\zeta(x_0) = 0$ and $\zeta'(x_0) =: \zeta'_0 > 0$, and we wish to integrate \eqref{eigenEasymp} backwards. 
For concreteness, let us say that $V(x) = 0$ for $x > x_0$, so $x_0$ corresponds to the edge of the ``cliff''.
Suppose that, in this cliff region, $(0,x_0)$, the potential is piecewise constant, so that in each interval where $k(x) = k$ is constant, $\zeta$ is a harmonic wave with period $l = \frac{2\pi}{k}$. The structure of $k(x)$ will follow a sequence of self-similar blocks, where the overall strength (i.e., value of $k$) doubles and the period halves from block to block. 
Moreover, each block consists of two stages, which we will describe by using as an example the first block, occupying the region $(\frac{x_0}{2}, x_0)$. Starting at $x_0$, let $k(x) = k_0$ for $3/4$ of its period, i.e., in the region $(x_0 - \frac{3}{4} \frac{2\pi}{k_0}, x_0)$. 
Then the strength is quadrupled so that $k(x) = 4k_0$ for $1/4$ of this new period, i.e., in the region $(x_0 - \frac{3}{4} \frac{2\pi}{k_0} - \frac{1}{4} \frac{2\pi}{4k_0},x_0 - \frac{3}{4} \frac{2\pi}{k_0})$. 
We wish to impose that this block ends at $\frac{x_0}{2}$, that is, 
\be\label{x0k0}
x_0 - \frac{3}{4} \frac{2\pi}{k_0} - \frac{1}{4} \frac{2\pi}{4k_0} = \frac{x_0}{2} \,\,\,\Longrightarrow\,\,\, k_0 x_0 = \frac{13\pi}{4}
\ee
This guarantees that the sequence of blocks terminates at $x=0$. 
At the end of the first stage of this block, $x=x_0 - \frac{3}{4} \frac{2\pi}{k_0}$, $\zeta$ will be at a local peak with value $\zeta_0 := \frac{\zeta_0'}{k_0}$; and at the end of the second stage (i.e., $x_1 := \frac{x_0}{2}$), it will have returned to zero value and with velocity $\zeta_1' := (4k_0) \zeta_0 = 4 \zeta_0'$. The second block, in the region $(\frac{x_0}{4}, \frac{x_0}{2})$, with duplicated frequencies $2k_0$ and $4(2k_0)$ for its two stages, thus starts with similar conditions as the previous block, but now with a quadrupled initial velocity $\zeta_1'$ (see Fig.~\ref{fig:potentialV}).
Explicitly, the cliff potential is given by
\be\label{Vexplic}
V(x) = \sum_{n\ge 0} 4^n f(2^n x)
\ee
where
\be
f(x) := \begin{cases}
    V_0 & \text{if } \frac{7x_0}{13} < x \le x_0\,, \\
    16V_0  & \text{if } \frac{x_0}{2} < x \le \frac{7x_0}{13}\,, \\
    0  & \text{otherwise.}
\end{cases}
\ee
with $V_0 := -\hbar^2k_0^2/2m$.

It is clear, by induction, that at the end of the $n$-th block, we will be sitting at 
\be
x_n := \frac{x_0}{2^n}
\ee
with $\zeta(x_n) = 0$ and $\zeta'_n := \zeta'(x_n) =  4^n\zeta_0'$.
Moreover, the peak value of $\zeta$ up to this point, which is attained in the $n$-th block, is
\be
\sup_{x \in (x_n, x_0)} |\zeta(x)| = \frac{\zeta_{n-1}'}{(2^{n-1}k_0)} = 2^{n-1} \zeta_0
\ee
Since from one block to the next the peak amplitude doubles, 
while the width of the block is halved, the norm within the block doubles. Hence the norm accumulated up to this point is
\be
\int_{x_n}^{x_0}\!dx\, |\zeta(x)|^2 = \sum_{m=0}^{n-1} 2^m \int_{x_0/2}^{x_0}\!dx |\zeta(x)|^2 = (2^n - 1) \frac{13\pi {\zeta'_0}^2}{16 k_0^3}
\ee
which diverges as $n\rightarrow \infty$ (i.e., $x_n \rightarrow 0$).
Therefore, $V(x)$ is in the limit point case at zero. Clearly, $V(x)$ is also limit point at infinity, since for $x>x_0$ the two solutions to \eqref{eigenEasymp} are spanned by $x$ and $1$, which have infinite near-infinity norm. Thus, $H$ is self-adjoint on its maximal domain, and the system is quantum-mechanically closed.

Finally, let us describe the other solution, $\varphi(x)$, obtained with initial values $\varphi(x_0) = \varphi_0$ and $\varphi'(x_0) = 0$. At $x_0-\frac{3}{4}\frac{2\pi}{k_0}$, $\varphi$ will be at zero value with velocity $\varphi'_0 := - k_0 \varphi_0$. Then, after the resistive work of $4k_0$, $\varphi$ will attain a peak value $-\frac{\varphi'_0}{(4k_0)} = \frac{\varphi_0}{4} =: \varphi_1$ at $x_1$. Note that, after each block, the peak value of $\varphi$ decreases by a factor of $4$, so $\lim_{x\rightarrow 0} \varphi(x) = 0$. Consequently, this solution is bounded near zero and the norm ``within the cliff'', $\int_{0}^{x_0}\!dx |\varphi(x)|^2$, is finite. Note also that, after each block, the maximum value of $\varphi'$ decreases by a factor of $2$, so $\lim_{x\rightarrow 0^+} \varphi'(x) = 0$. Since this solution $\varphi(x)$ is real, the probability current is automatically zero, confirming that there is no ``leakage'' of the wavefunction through zero.

\section{Spectrum and scattering}
\label{sec:dynamics}

In this section we study the dynamical properties of the impassable cliff potential in more depth, by deriving the energy eigenstates, discussing the spectrum and computing the phase shift for a scattering state reflecting from the cliff.

\subsection{Energy eigenstates}
\label{subsec:eigen}

Let us consider the eigenvalue problem
\be\label{eigenE}
- \frac{\hbar^2}{2m} \frac{d^2 \psi}{dx^2}(x) + V(x) \psi(x) = E\psi(x)
\ee
with $E \in \bb R$, a priori.
If we start at some point $x > 0$ with generic initial conditions, $\psi(x)$ and $\psi'(x)$, and integrate this equation towards $x=0$, the solution will generally converge asymptotically to a superposition of $\zeta$ and $\varphi$. As $\zeta$ does not belong to the Hilbert space, we must restrict the initial conditions so that it asymptotically converges to something proportional to $\varphi$ as $x\to 0^+$. 

It is useful to convert \eqref{eigenE} into a system of first-order differential equations,
\be
\frac{d}{dx}\begin{pmatrix} \psi(x) \\ \psi'(x) \end{pmatrix} = \begin{pmatrix} 0 & 1 \\ -k(x)^2 \!-\! \scr E & 0 \end{pmatrix} \begin{pmatrix} \psi(x) \\ \psi'(x) \end{pmatrix}
\ee
where
\be
\scr E := \frac{2mE}{\hbar^2}
\ee
The solution, integrating down to $\epsilon >0$, is
\be\label{eigenkernel}
\begin{pmatrix} \psi(\epsilon) \\ \psi'(\epsilon) \end{pmatrix} = W(x, \epsilon; \scr E) \begin{pmatrix} \psi(x) \\ \psi'(x) \end{pmatrix}
\ee
with
\be
W(x, \epsilon; \scr E) := \pexp \left[ \int_x^\epsilon\!dx' \begin{pmatrix} 0 & 1 \\ -k(x')^2 \!-\! \scr E & 0 \end{pmatrix} \right] 
\ee
where $\pexp$ denotes the path-ordered exponential (in which greater $x'$ appears to the right).
To obtain a solution behaving like $\varphi$ near zero, we must impose that the left-hand side of \eqref{eigenkernel} vanishes in the limit $\epsilon \rightarrow 0$.

\begin{figure}[h!]  
\centering
\includegraphics[scale = 0.68]{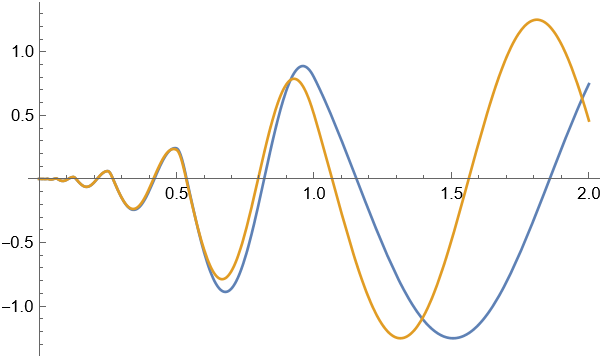}
\caption{\small The (generalized) eigenfunctions $\psi(x)$, for $0<x<2$, for energies $E = 10$ (blue) and $E=20$ (orange). Here $x_0 = 1$, $m=1$ and $\hbar = 1$. Note that as $x\rightarrow 0^+$ they asymptotically converge to $\varphi(x)$, and for $x \ge x_0$ they behave as in \eqref{psiharmonic}.}
\label{fig:psipositive}
\end{figure}

However, we cannot simply set $\epsilon = 0$ above since $\lim_{\epsilon\rightarrow 0} W(x, \epsilon; \scr E)$ is ill-defined. 
Rather, taking $\epsilon = x_n$, in the limit where the integer $n\rightarrow \infty$, we can impose that the left-hand side of \eqref{eigenkernel} is proportional to $\left(\begin{smallmatrix} 1 \\ 0 \end{smallmatrix}\right)$, since $\varphi$ is precisely defined by having zero derivative at the beginning of each block. 
The condition is thus, 
\be
W_{21}(x, x_n; \scr E) \psi(x) + W_{22}(x, x_n; \scr E) \psi'(x) = 0
\ee
as $n$ goes to infinity. If $n \gg 1$, the energy can be neglected in the $n$-th block, sitting between $x_{n-1}$ and $x_n$, whose contribution to the path-ordered exponential is therefore
\be\label{transfermatblock}
\pexp \left[ \int_{x_{n-1}}^{x_n}\!dx' \begin{pmatrix} 0 & 1 \\ -k(x')^2 \!-\! \scr E & 0 \end{pmatrix} \right] \approx \begin{pmatrix} \sfrac{1}{4} & 0 \\ 0 & 4 \end{pmatrix}
\ee
Consequently, as $n$ grows large, $W_{21}(x, x_n; \scr E)$ and $W_{22}(x, x_n; \scr E)$ will grow as $4^n$. 
By properly rescaling these matrix elements with $4^{-n}$, we can define the limiting vector
\be
\fr W(x;\scr E) := \lim_{n\rightarrow\infty} 4^{-n}  \begin{pmatrix}  W_{22}(x,x_n;\scr E) \\ - W_{21}(x,x_n;\scr E) \end{pmatrix}
\ee
Then, any valid initial data at $x$ for an eigenfunction with energy $E = \hbar^2 \scr E/2m$ must satisfy 
\be\label{Psicd}
\begin{pmatrix} \psi(x) \\ \psi'(x) \end{pmatrix} \propto \fr W(x;\scr E)
\ee
If $E$ belongs to the spectrum, the corresponding eigenfunction is given, for $x>x_0$, by
\be\label{psiharmonic}
\psi(x) = \fr W_1(x_0;\scr E) \cos\big(\sqrt{\scr E}(x-x_0)\big)\, +\frac{\fr W_2(x_0;\scr E)}{\sqrt{\scr E}} \sin\big(\sqrt{\scr E}(x-x_0)\big) 
\ee
See Fig.~\ref{fig:psipositive} for the plot of two eigenfunctions with positive energy (more precisely, those are ``generalized eigenfunctions'' since they are not square-integrable near infinity, even though all positive energies belong to the spectrum, as we will see in the next subsection).

\subsection{Energy spectrum}
\label{subsec:spectrum}

We now wish to determine which energies belong to the spectrum of the Hamiltonian.
As the behavior of the eigenfunctions is under control as $x\to 0^+$, given condition \eqref{Psicd}, it only remains to ensure that those eigenfunctions are well-behaved as $x\to\infty$.
From the general solution for $x>x_0$, given in \eqref{psiharmonic}, we see that the eigenfunctions are harmonic functions near infinity for any positive energy. While those are not square integrable, and thus do not belong to the Hilbert space, they still make sense as ``generalized eigenfunctions'' associated with the continuous spectrum.
Thus, like in the case of a free particle ($V=0$), every $E \ge 0$ belongs to the spectrum of the Hamiltonian.\footnote{In particular, for any $E$ in the spectrum of $H$, there is a sequence of (normalizable) states $\psi_n$ such that ${\lim_{n\rightarrow\infty} |\!| (H - E) \psi_n |\!|/|\!|\psi_n|\!|} = 0$. 
The harmonic functions are sufficiently ``mild'' so as to allow this limiting interpretation.
Contrastingly, an eigenfunction behaving like $\zeta$ near zero or like a growing exponential near infinity do not admit such a limiting interpretation, and hence must be excluded altogether.\label{limiteigen}}

Since the potential is negative, and in fact unbounded from below, it is not surprising that there are also negative energies.
The negative portion of the spectrum is, however, discrete. The reason is that the negative energy solutions given in \eqref{psiharmonic} will generally consist of one diverging and one converging exponential for $x\ge x_0$, and only the latter is allowed (see footnote~\ref{limiteigen}). It is necessary to ``fine tune'' $E$ so that, starting from $x_0$, the solution of \eqref{eigenE} converges to ${\sim\varphi}$ as $x\rightarrow 0^+$ and to
${\sim e^{-\sqrt{-\scr E}\,x}}$ for $x \ge x_0$. The condition is thus that $\psi'(x_0) = -\sqrt{-\scr E} \psi(x_0)$, which from \eqref{Psicd} corresponds to
\be\label{negativeE}
\Gamma(x_0;\scr E) := \fr W_2(x_0;\scr E) + \fr W_1(x_0;\scr E)\sqrt{-\scr E} = 0
\ee

Many interesting properties about the negative spectrum can be studied analytically. 
First, note that by defining a dimensionless position variable $\sigma := x/x_0$, we can rewrite equation \eqref{eigenE} as 
\be\label{eigenE2}
 \frac{d^2 \Psi}{d\sigma^2}(\sigma) = - \left( \kappa(\sigma)^2 + \lambda \right) \Psi(\sigma)
\ee
where $\Psi(\sigma) := \psi(x_0\sigma)$, $\kappa(\sigma) := x_0 k(x_0 \sigma)$ and
\be
\lambda := \frac{2mE x_0^2}{\hbar^2} = \scr E x_0^2
\ee
In this new form, the equation is completely dimensionless and, due to relation \eqref{x0k0}, $\kappa(\sigma)$ is independent of the physical parameters (i.e., $m$, $x_0$ and $\hbar$). We refer to the eigenvalues $\lambda$ as the \emph{dimensionless eigenvalues}, which are independent of the physical parameters.
From a numerical computation we find that the first few elements of the negative part of the spectrum of $H$ are
\ba
\text{\emph{Spec}}^-(H) =  \frac{\hbar^2}{2mx_0^2}\{&-72.6416, -210.342, -715.831, \no
&-841.391, -2863.33, -3365.56, \no
&-11453.3, -13462.3, -45813.2, \no
&-53849., -183253.,\,\ldots\} \label{Speclambda}
\ea

\begin{figure}[h!]  
\centering
\includegraphics[scale = 0.68]{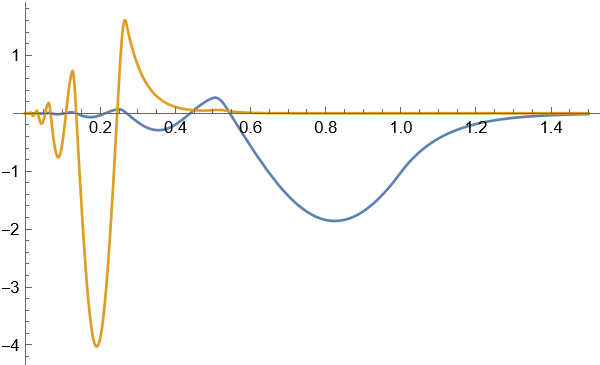}
\caption{\small The (normalized) eigenfunctions $\psi(x)$, for $0<x<1.5$, for energies $E = -72.6416$ (blue) and $E=-841.391$ (orange). Here $x_0 = 1$, $m=1/2$ and $\hbar = 1$ (so that $E = \lambda$). Note how the eigenstate with large negative energy (fourth root in \eqref{Speclambda}) is localized deeply within the cliff, with only $3.174\times 10^{-16}$ probability lying outside ($x>x_0$).}
\label{fig:psinegative}
\end{figure}

Let us show that if $E<0$ is large and belongs to the spectrum of $H$, then there is another eigenvalue very close to $E/4$.
First, note that for large negative $E$ the eigenfunction is concentrated deeply inside the cliff, near $x=0$, as expected for bound states (see Fig.~\ref{fig:psinegative}).
So the eigenfunction would barely ``feel'' if the first block of the potential were deleted. Consequently, if $E$ is an eigenvalue for a cliff with edge at $x_0$, then there should exist an eigenvalue $E_1 \approx E$ for a cliff with edge at $x_1 = x_0/2$. 
It follows that $\lambda_1 = \frac{2mE_1 x_1^2}{\hbar^2}$ is one of the dimensionless eigenvalues, and consequently 
\be
\frac{\hbar^2}{2mx_0^2} \lambda_1 =  \frac{E_1}{4} \approx \frac{E}{4}
\ee
is in the spectrum for a cliff with edge at $x_0$.

The argument above can be formalized as follows. According to \eqref{Vexplic}, the potential with the first block ($n=0$) deleted is given by $V_1(x) = V(x) - f(x)$. The Hamiltonian for a cliff with edge at $x_1 = x_0/2$ is then $H_1 = H - f$. 
Since $H_1$ is self-adjoint, for any normalized wavefunction $\psi$ (in its domain) we have
\be
\text{Dist}[E, \text{Spec}(H_1)] \le |\!|(H_1 - E)\psi |\!|
\ee
Taking $\psi$ to be a (normalized) eigenfunction of $H$ with eigenvalue $E$, we get
\be
\text{Dist}[E, \text{Spec}(H_1)] \le |\!|(H_1 - H)\psi |\!| = |\!|f\psi |\!|
\ee
As $f$ is supported in $(x_1, x_0)$, we just need to estimate the portion of $\psi$ in this region.
For $x \gtrsim x_*$, where $x_*$ is the point at which $k(x_*)^2 \sim -\scr E$, the (normalized) eigenfunction behaves roughly like an exponential,
\be
\psi(x) \sim (-\scr E)^{1/4} e^{-\sqrt{-\scr E} (x-x_*)}
\ee
When $-E \gg -V_0$, we have $x_* \ll x_0$, $\sqrt{-\scr E} x_0 \gg 1$, so
\be
|\!|f\psi|\!| \sim |V_0| e^{-\sqrt{-\scr E} x_0/2} \ll |V_0|
\ee
confirming that $H_1$ has an element in its spectrum not farther from $E$ than roughly this small bound.

A consequence of this is that for any large $E<0$ in the spectrum, $E$ can be divided by $4$ repeatedly until eventually it will come very close to one of the values in \eqref{Speclambda}. In fact, the agreement is already good from the fourth root, $(-841.391)/4 = -210.348 \approx -210.342$. 
A similar argument also holds in the other direction, i.e., if $E$ is a large negative eigenvalue, then there is another eigenvalue near $4E$ (which is derived by inserting a new block between $x_0$ and $2x_0$). 
It follows that the first few eigenvalues (given in \eqref{Speclambda}) approximately determine all others, and that the spectrum is unbounded from below.

\subsection{Phase shift}
\label{subsec:phase}

We have seen that the spectrum consists of all non-negative energies, plus an infinite discrete negative part. The positive energy eigenfunctions behave as harmonic functions outside the cliff region, and are thus scattering states. The negative energy states are mostly concentrated within the cliff region, decaying exponentially outside, and are thus bound states. Here we compute the phase shift associated with an incoming state from infinity reflecting from the cliff, which is an important parameter in scattering theory.

The phase shift is already encoded in the vector $\fr W(x_0;\scr E)$, whose components determine the coefficients of the sine and cosine modes in the region outside the cliff, as given in \eqref{psiharmonic}. 
Choosing the edge of the cliff as the reference point, we simply need to reexpress the sine and cosine in terms of left and right moving waves, respectively $e^{-i\sqrt{\scr E}(x-x_0)}$ and $e^{i\sqrt{\scr E}(x- x_0)}$. We get
\be
\psi(x) \propto e^{-i\sqrt{\scr E}(x-x_0)} + \frac{\sqrt{\scr E}\,\fr W_1(x_0;\scr E) - i \fr W_2(x_0;\scr E)}{\sqrt{\scr E}\,\fr W_1(x_0;\scr E) + i \fr W_2(x_0;\scr E)} e^{i\sqrt{\scr E}(x-x_0)}
\ee
Notice that the coefficient of $e^{i\sqrt{\scr E}(x-x_0)}$ (i.e., the reflection factor) has unit norm for every energy, as expected in a perfect reflection. 
Defining
\be
\vartheta(E) := 2\text{Arg}\big[\sqrt{\scr E}\,\fr W_1(x_0;\scr E) + i \fr W_2(x_0;\scr E)\big]
\ee
we obtain 
\be
\psi(x) \propto e^{-i\sqrt{\scr E}(x-x_0)} + e^{-i\vartheta(E)} e^{i\sqrt{\scr E}(x-x_0)}
\ee
so $\vartheta(E)$ is the phase shift at energy $E$.

\section{Conclusion}
\label{sec:conc}

We have constructed a potential that is everywhere finite
on the half-line and, despite being unbounded from below and nowhere positive,
completely prevents the particle from escaping through the edge. 
We have clarified the importance of specifying the domain of the Hamiltonian in order to determine whether it is
self-adjoint, and consequently whether time evolution is well-defined and unitary. 
The crucial point is that a formal quantization of the classical Hamiltonian does not automatically select a domain. We argued that in the absence of additional physical data (including boundary conditions), it is most natural to define the Hamiltonian operator on its maximal domain.
Self-adjointness then becomes a property of the quantized dynamics rather than a condition imposed by hand.
We emphasize that the mechanism whereby the ``impassable cliff'' manages to confine the particle in the half-line is purely dynamical, following strictly from quantization, and does not require any boundary conditions or structure extrinsic to the configuration space.

Beyond establishing dynamical closure, we analyzed the energy eigenstates of the impassable cliff. The non-negative spectrum is continuous and describes scattering states that are perfectly reflected with an energy-dependent phase shift, while the negative spectrum consists of discrete bound states, is unbounded from below, and becomes approximately invariant under $E\to 4E$ at large negative energies.

It is worth mentioning that, while our potential provides a particular example constructed as a series of piecewise constant steps, a much larger class of potentials unbounded from below, including smooth ones, could similarly be naturally self-adjoint on the half-line. In particular, it is only necessary that integrating the time-independent Schr\"odinger equation towards $x=0$, starting from arbitrary initial data at some $x$, produce at least one solution with diverging $L^2$ norm near zero. For potentials constructed as self-similar blocks,
that occurs whenever the repeated ``transfer matrices'' (introduced in Sec.~\ref{subsec:eigen}) drive one solution to sufficiently rapid growth that its near-zero norm diverges, as happens in \eqref{transfermatblock}.

Related behavior occurs in familiar examples where a family of negative potentials become opaque in the singular infinite-depth limit \cite{messiah2014quantum}. For instance, an attractive square well can approach perfect reflection for suitably tuned widths as its depth tends to infinity. In such examples, however, the potential does not converge to a limiting function, even distributionally, so the Schr\"odinger equation is not well-defined after the limit is taken. 
Moreover, when such constructions are used to confine a particle to the half-line, the confining mechanism takes place outside the intended configuration space, upon extending the half-line into the whole line (e.g., a family of cliff potentials with $V(x) = 0$ for $x>0$ and $V(x) = V_0 \to -\infty$ for $x < 0$).
Another familiar phenomenon with some superficial resemblance is localization in condensed-matter physics, where suitably structured or disordered potentials can strongly suppress propagation \cite{anderson1958absence}. 
There, however, the confinement is approximate in the sense that localized states generally retain spatially decaying tails.
The impassable cliff is fundamentally different from these two classes of examples: it is a single, well-defined cliff potential on $\bb R^+$ that generates exact confinement entirely within the configuration space, defining a dynamically closed quantum system.

%\newpage 
\vskip 3em
\section*{Acknowledgements}

I especially thank Ted Jacobson for valuable discussions and insights,
and providing extensive feedback on the draft of this paper.
I thank also Pranav Pulakkat, Fritz Gesztesy, Gerald Teschl and Barbara \v{S}oda for useful conversations.
This research was supported by the National Science Foundation under Grant PHY-2309634 and Perimeter Institute for Theoretical
Physics. Research at Perimeter Institute is supported
in part by the Government of Canada through
the Department of Innovation, Science and Economic Development
and by the Province of Ontario through the
Ministry of Colleges and Universities.

\bibliographystyle{ieeetr}
\bibliography{BibRAS}

\begin{thebibliography}{10}

\bibitem{rauch1973two}
J.~Rauch and M.~Reed, ``Two examples illustrating the differences between classical and quantum mechanics,'' {\em Communications in Mathematical Physics}, vol.~29, no.~2, pp.~105--111, 1973.

\bibitem{reed1975ii}
M.~Reed and B.~Simon, {\em II: Fourier analysis, self-adjointness}, vol.~2.
\newblock Elsevier, 1975.

\bibitem{Vilenkin_1994}
A.~Vilenkin, ``Approaches to quantum cosmology,'' {\em Physical Review D}, vol.~50, p.~2581–2594, Aug. 1994.

\bibitem{pinto2013quantum}
N.~Pinto-Neto and J.~Fabris, ``{Quantum cosmology from the de Broglie-Bohm perspective},'' {\em Classical and Quantum Gravity}, vol.~30, no.~14, p.~143001, 2013.

\bibitem{pinto2012wheeler}
N.~Pinto-Neto, F.~Falciano, R.~Pereira, and E.~S. Santini, ``{Wheeler-DeWitt quantization can solve the singularity problem},'' {\em Physical Review D}, vol.~86, no.~6, p.~063504, 2012.

\bibitem{Kim_1997}
S.~P. Kim, ``Quantum potential and cosmological singularities,'' {\em Physics Letters A}, vol.~236, p.~11–15, Dec. 1997.

\bibitem{harlow2020factorization}
D.~Harlow and D.~Jafferis, ``{The factorization problem in Jackiw-Teitelboim gravity},'' {\em Journal of High Energy Physics}, vol.~2020, no.~2, pp.~1--32, 2020.

\bibitem{hall2013quantum}
B.~C. Hall, {\em Quantum theory for mathematicians}.
\newblock Springer, 2013.

\bibitem{bonneau2001self}
G.~Bonneau, J.~Faraut, and G.~Valent, ``Self-adjoint extensions of operators and the teaching of quantum mechanics,'' {\em American Journal of physics}, vol.~69, no.~3, pp.~322--331, 2001.

\bibitem{araujo2004operator}
V.~S. Araujo, F.~A.~B. Coutinho, and J.~Fernando~Perez, ``Operator domains and self-adjoint operators,'' {\em American Journal of Physics}, vol.~72, no.~2, pp.~203--213, 2004.

\bibitem{Weyl1910}
H.~Weyl, ``Über gewöhnliche differentialgleichungen mit singularitäten und die zugehörigen entwicklungen willkürlicher funktionen. (mit 1 figur im text),'' {\em Mathematische Annalen}, vol.~68, pp.~220--269, 1910.

\bibitem{zettl2005sturm}
A.~Zettl, {\em Sturm-liouville theory}.
\newblock No.~121, American Mathematical Soc., 2005.

\bibitem{messiah2014quantum}
A.~Messiah, {\em Quantum mechanics}.
\newblock Courier Corporation, 2014.

\bibitem{anderson1958absence}
P.~W. Anderson, ``Absence of diffusion in certain random lattices,'' {\em Physical review}, vol.~109, no.~5, pp.~1492--1505, 1958.

\end{thebibliography}

\end{document}